\documentclass[
aps,nofootinbib,showpacs,showkeys,preprint
] {revtex4}

\usepackage{epsf,epsfig,subfigure,
graphicx,amsmath,amssymb}
\usepackage{color}

\begin{document}

\begin{flushright}
{\tt 
~~~~ \quad
\\}
\end{flushright}

\title{\Huge\bf Spectral Index and Non-Gaussianity  
\\ in Supersymmetric Hybrid Inflation}

\author{\large Bumseok Kyae\email{bkyae@pusan.ac.kr} } 

\affiliation{
Department of Physics, Pusan National University, Busan
609-735, Korea
 \\~~}

\vspace{0.7cm}

\begin{abstract}

We consider a supersymmetric hybrid inflation model with two
inflaton fields. The superpotential during inflation is dominated
by $W=(\kappa S+\kappa' S')M^2$, where $S$, $S'$ are inflatons carrying the same U(1)$_R$ charge, $\kappa$, $\kappa'$
are dimensionless couplings, and $M$ ($\sim 10^{15-16}$ GeV) is a
dimensionful parameter associated with a symmetry breaking scale.
One light mass eigenstate drives inflation, while the other
heavier mass eigenstate is stuck to the origin. The smallness of
the lighter inflaton mass for the scalar spectral index
$n_s\approx 0.96$, which is the center value of WMAP7, can be
controlled by the ratio $\kappa'/\kappa$ through the supergravity corrections. We also discuss the possibility of the two field inflation and large non-Gaussianity in this setup. 

\end{abstract}

\pacs{98.80.Cq, 12.60.Jv, 04.65.+e}

\keywords{Inflation, Supergravity, Spectral index, Non-Gaussianity} 

\maketitle

\section{Introduction}

Although inflation seems to be inevitable in cosmology to resolve
the homogeneous and flatness problems, it is highly nontrivial to
realize the idea in the scalar field theory framework
\cite{review}. It is because a small scalar mass is perturbatively
unstable, and a heavy inflaton mass term destroys the slow-roll
conditions needed for sufficient e-folds ($\gtrsim 50$--60).
Supersymmetry (SUSY) is helpful for keeping the smallness of the
inflaton mass against quantum corrections, but it is just up to the
Hubble scale. Supergravity (SUGRA) correction in de Sitter space
usually induces a Hubble scale mass term of the inflaton at tree
level even with the minimal K${\rm \ddot{a}}$hler potential,
unless the model is carefully constructed. It is called the
``$\eta$ problem.''

One of the promising models, which potentially avoids the $\eta$
problem, is the SUSY hybrid inflation model
\cite{FtermInf2,FtermInf}.  In that model, the superpotential is
dominated by $W=\kappa SM^2$ during the inflation era, where $S$
denotes the inflaton superfield, and  $\kappa$ and $M$ are
dimensionless and dimensionful parameters, respectively. $M$ turns
out to be associated with a symmetry breaking scale. With this
superpotential, the SUGRA correction does not induce the dangerous
Hubble scale inflaton mass term ($3H^2|S|^2$) in the scalar
potential, if the K${\rm \ddot{a}}$hler potential is given by the
minimal form ($K=|S|^2$) \cite{FtermInf2,SUGRAcorr}: such a mass
term is accidentally canceled out at tree level in this model.
However, a quartic term with the dimensionless coefficient of
order unity in the K${\rm \ddot{a}}$hler potential, $K\supset
c_1|S|^4/4M_P^2$, where $M_P$ is the reduced Planck mass ($\approx
2.4\times 10^{18}$ GeV), generates the unwanted inflaton mass term
in the scalar potential. Actually, only the quartic term in the
K${\rm \ddot{a}}$hler potential is dangerous, while higher order
terms with coefficients of order unity are all harmless, because
$|S|\lesssim M_P$. Therefore, only the coefficient $c_1$ should be
assumed to be adequately suppressed ($\lesssim 10^{-2}$). This
naive assumption needs to be justified by a UV theory or a quantum
gravity theory in the future.

A remarkable feature in the SUSY hybrid inflation model is that
the CMB anisotropy $\delta T/T$ is proportional to $(M/M_P)^2$
\cite{FtermInf}. Thus, the observational data of $\delta T/T\sim
10^{-5}$ \cite{WMAP7} determine the spontaneous symmetry breaking
scale: $M\approx 10^{15-16}$ GeV, which is tantalizingly close to
the scale of SUSY grand unified theory (GUT) \cite{FtermInf}.  As
a result, the SUSY hybrid inflation model can be embedded in the
models of SUSY GUT. Indeed, this idea has been combined with the
particle physics models of
SU(3)$_c\times$SU(2)$_L\times$SU(2)$_R\times$U(1)$_{B-L}$
\cite{3221}, SU(4)$_c\times$SU(2)$_L\times$SU(2)$_R$ \cite{422},
SU(5)$\times$U(1)$_X$ \cite{FlippedSU(5)}, and SO(10)
\cite{SO(10)}.  In those models, $M$ is interpreted as the
U(1)$_{B-L}$ breaking scale.

In the SUSY hybrid inflation model, the scalar spectral index is
predicted:
\begin{eqnarray} \label{spectral}
n_s\approx 1+2\eta\approx 1-\frac{1}{N_e}\approx 0.98 ,
\end{eqnarray}
where $\eta$ ($\equiv M_P^2V''/V$) is the slow-roll parameter, and
$N_e$ denotes the e-folding number ($=50$--$60$). On the other
hand, the recent WMAP 7-year (WMAP7) observation result on the
scalar spectral index is $n_s=0.96^{+0.014}_{-0.013}$
\cite{WMAP7}. Thus, the prediction of the SUSY hybrid inflation
model, $n_s\approx 0.98$ is quite deviated from the center value
of the WMAP7 result. Indeed, unless relatively larger SUGRA
corrections are included or the model is much ameliorated, the
deviation is not easily overcome.

Ref. \cite{hilltop} shows that the relatively smaller spectral
index can be achieved, particularly if inflation takes place near
the local maximum of the scalar potential. For obtaining such a
local maximum of the inflaton's scalar potential, a negative mass
term of the inflaton is needed. In the SUSY hybrid inflationary
model, it can be supported from the non-minimal K${\rm
\ddot{a}}$hler potential.

The literatures attempted to explain the center value of $n_s$
($\approx 0.96$) by considering a small (but relatively larger)
quartic \cite{hilltop,nonminimal-K} term and/or a more higher
order term \cite{quartic-term} in the K${\rm \ddot{a}}$hler
potential, or a soft SUSY breaking ``A-term'' \cite{a-term} in the
scalar potential. In this paper, we will point out that
%
%
the {\it superpotential}, which is quantum mechanically
controllable, also can play an essential role in explaining
$n_s\approx 0.96$.
We will discuss also the possibility of the two field inflation and large non-Gaussianity in this setup. 


\section{``Hilltop Inflation''}

The SUSY hybrid inflation model is defined as the following
superpotential \cite{FtermInf2,FtermInf}:
\begin{eqnarray} \label{original}
W=\kappa S(M^2-\psi\overline{\psi}) ,
\end{eqnarray}
where $\psi$ and $\overline{\psi}$ are a conjugate pair of
superfields carrying gauge and/or global charges. At the SUSY
minimum, $S=0$ and $|\psi|=|\overline{\psi}|=M$ by including the
D-term potential, breaking a symmetry by the non-zero vacuum
expectation values (VEVs) of $\psi$ and $\overline{\psi}$.
Inflation starts when the inflaton $S$ is much deviated from the
minimum, $S\gtrsim M$. Then, the complex scalars, $\psi$ and
$\overline{\psi}$ obtain heavy masses, by which
$\psi=\overline{\psi}=0$ during inflation. It is the quasi-stable
point. Thus, the superpotential is dominated by $W=\kappa SM^2$
during inflation. It provides a positive constant vacuum energy
density $\kappa^2 M^4$, which gives rise to inflation. As
mentioned above, the superpotential $W=\kappa SM^2$ and the
minimal K${\rm \ddot{a}}$hler potential do not raise the ``$\eta$
problem.'' Since the higher order terms of the singlet $S$ in the
superpotential destroys the slow-roll conditions, they should be
forbidden by introducing the U(1)$_R$ symmetry.\footnote{U(1)$_R$
is eventually broken e.g. by SUSY breaking effect or instanton
effect. If U(1)$_R$ is broken at the hidden sector, the U(1)$_R$
axion cannot be the QCD axion.} The triggering condition for
inflation, $S\gtrsim M$ would be possible, if the universe was hot
enough before inflation was initiated.

Because of the positive vacuum energy, SUSY is broken and so the
constant scalar potential is quantum mechanically corrected.
Neglecting the SUGRA corrections, thus, the scalar potential is
given by \cite{FtermInf}
\begin{eqnarray} \label{CWpot}
V\approx \kappa^2M^4\left(1+\frac{\kappa^2}{16\pi^2}~{\rm
log}\frac{\kappa^2|S|^2}{\Lambda^2}\right) ,
\end{eqnarray}
where the logarithmic term denotes the quantum correction when
$S\gtrsim M$, and $\Lambda$ means the renormalization scale. It
makes a small slope in the potential, leading the inflaton to the
SUSY minimum. As shown in Eq.~(\ref{spectral}), however, the
scalar potential Eq.~(\ref{CWpot}) yields $n_s\approx 0.98$,
unless it is somehow modified.

Let us consider the following form of the modified inflaton
potential:
\begin{eqnarray} \label{toypot}
V=\mu^4\left(1+\alpha ~{\rm
log}~\varphi+\frac{\delta}{2}\varphi^2\right) ,
\end{eqnarray}
where $\mu^4$ is the positive vacuum energy density leading to
inflation. The dimensionless field $\varphi$ denotes an inflaton
scalar defined as $S/M_P$. The logarithmic term arises from the
quantum correction caused by SUSY breaking \cite{FtermInf}.
Comparison with Eq.~(\ref{CWpot}) yields the relations,
$\mu^4\approx\kappa^2M^4$ and $\alpha\approx\kappa^2/8\pi^2$. In
Eq.~(\ref{toypot}), the inflaton's mass term is introduced:
$V\supset (\delta/2)\mu^4\varphi^2=(3\delta/2)H^2S^2$, where $H$
($=\sqrt{\mu^4/3M_P^2}$) is the Hubble constant during inflation.
For successful inflation, thus, the dimensionless coupling
$\delta$ should be small enough, $|\delta|\ll 1$.

The slow-roll parameter $\epsilon$ is still much smaller than
$|\eta|$. It is basically because $\varphi$ is assumed to be
smaller than the unity. In the presence of the mass term in
Eq.~(\ref{toypot}), the expressions of the e-folding number ``$N_e$'' and
``$\eta$'' are given by \cite{hilltop}
\begin{eqnarray} \label{N,eta}
N_e=\frac{1}{2\delta}~{\rm
log}\left(1+\frac{\delta}{\alpha}\varphi^2\right) ~, \quad {\rm
and} \quad\quad \eta=\delta \times\frac{e^{2\delta
N_e}-2}{e^{2\delta N_e}-1} ~.
\end{eqnarray}
We note that in the limit of $\delta\rightarrow 0$, the expression
for $N_e$ and $\eta$ become $\varphi^2/2\alpha$ and $-1/2N_e$,
respectively, which are the expressions given in the original form
of the SUSY hybrid inflation model. In Eq.~(\ref{N,eta}), the
limit $\alpha\rightarrow 0$ does not make sense.  It means that
the logarithmic quantum correction makes an important contribution
to $N_e$.

With the help of the inflaton mass term, the scalar spectral index
can be compatible with the center value of WMAP7:
\begin{eqnarray} \label{ns}
n_s ~ \approx ~ 1+2\eta ~ \approx ~ 0.96 \quad\quad{\rm for}\quad
\frac{\delta}{2} = -3.5\times 10^{-3} .
\end{eqnarray}
Here we set $N_e=55$, but $n_s$ is quite insensitive to large
$N_e$s. Since the sign of the quadratic mass term in
Eq.~(\ref{toypot}) is negative, the potential is convex-upward.
If the inflaton starts at a point of $V'>0$ or
$\alpha+\delta\cdot\varphi^2>0$, the inflaton can roll down
eventually to the origin. It is fulfilled for $\kappa\gtrsim
5\times 10^{-2}$ ($5\times 10^{-3}$) and $\varphi\sim 0.1$
($0.01$). Actually, inflation would take place near the local
maximum, ``hilltop'' \cite{hilltop}, unless $\varphi \ll 1$.

The curvature perturbation is estimated as
\begin{eqnarray} \label{powerSP}
{\cal P}^{1/2}_{\cal R}= \frac{1}{\sqrt{12}\pi M_{P}^3}~
\frac{V^{3/2}}{V'} \approx \left(\frac{M}{M_{P}}\right)^2
\sqrt{\frac{2|1-e^{2\delta N_e}|}{3 |\delta|~e^{4\delta N_e}}} ,
\end{eqnarray}
where we set $\mu^2=\kappa M^2$.
For $N_e=55$, $\delta=-7.0\times 10^{-3}$, and ${\cal
P}^{1/2}_{\cal R}\approx 4.93\times 10^{-5}$ \cite{WMAP7}, $M$ is
approximately $4.3\times 10^{15}$ GeV, which is slightly lower
than that in the case of $\delta=0$ ($5.7\times 10^{15}$ GeV).
If $10^{12}$ GeV $\lesssim M\lesssim 10^{15}$ GeV, the curvature
perturbation should be supplemented by another scalar field,
``curvaton'' \cite{curvaton}. Then, inflation does not have to occur
near the local maximum, since the room between $M\lesssim S$ and
$S\ll M_P$ (or $\varphi \ll 1$) can be much larger. If $M\ll
10^{15}$ GeV, however, the inflationary scenario cannot be
embedded in a SUSY GUT model any more.
With the scalar potential Eq.~(\ref{toypot}), the fraction of the
tensor perturbation is unlikely to be detectable in the near
future.


\section{Twinflation}

The negative small inflaton mass squared in Eq.~(\ref{toypot}) can
be supported from a small quartic term of the K${\rm
\ddot{a}}$hler potential \cite{hilltop,nonminimal-K}, but we will
explore another possibility.
Let us introduce one more inflaton $S'$.  It carries the same
quantum number as $S$, but has a mass different from that of $S$.
Actually, no intrinsic reason why only one inflaton field $S$
should exist in nature can be found. In the presence of the twin
inflanton fields $\{S,S'\}$, and two pairs of the waterfall fields
$\{\psi_1,~\overline{\psi}_1\}$, $\{\psi_2,~\overline{\psi}_2\}$,
the general superpotential takes the following form:
\begin{eqnarray} \label{superPot}
W= S\left(\kappa_1M^2-\kappa_1\psi_1\overline{\psi}_1
-\kappa_2\psi_2\overline{\psi}_2\right)
+S^\prime\left(\kappa_2^\prime M^{\prime
2}-\kappa_1^\prime\psi_1\overline{\psi}_1
-\kappa_2^\prime\psi_2\overline{\psi}_2 \right) ,
\end{eqnarray}
where we assign the U(1)$_R$ charges of 2 (0) to $S$ and
$S^\prime$ ($\psi_{1,2}$ and $\overline{\psi}_{1,2}$) such that
the higher power terms of $S$ and $S^\prime$ are forbidden.\footnote{Since the U(1)$_R$ symmetry is unique in $N=1$ SUSY theory, $S^\prime$ should carry the exactly same charge with $S$. On the other hand, $\psi_2$ and $\overline{\psi}_2$ can carry the charges different from those of $\psi_1$ and $\overline{\psi}_1$, and respect a symmetry different from that $\psi_1$ and $\overline{\psi}_1$ respect. 
Such a symmetry could be anomaly-free. The issue associated with topological defects by  $\langle\psi_1\overline{\psi}_1\rangle$, $\langle\psi_2\overline{\psi}_2\rangle$ can be addressed by considering the higher power terms, $S^{(\prime)}(\psi_1\overline{\psi}_1)^2$ and $S^{(\prime)}(\psi_2\overline{\psi}_2)^2$, etc. in the superpotential, leaving intact the original inflationary scenario \cite{422}. 
}
The different coupling constants $\kappa_{1,2}$ and
$\kappa_{1,2}'$ distinguish $S$ and $S'$.  We assume that
$\kappa_{1,2}^{(\prime)}$ and $M^{(\prime) 2}$ are real quantities
for simplicity.  At the SUSY minimum, $\psi_{1,2}$ and
$\overline{\psi}_{1,2}$ get heavy masses as well as the VEVs,
satisfying $\psi_1^*=\overline{\psi}_1$ ($\neq 0$),
$\psi_2^*=\overline{\psi}_2$ ($\neq 0$), and
$\kappa_1M^2-\kappa_1\psi_1\overline{\psi}_1
-\kappa_2\psi_2\overline{\psi}_2=\kappa_2^\prime M^{\prime
2}-\kappa_1^\prime\psi_1\overline{\psi}_1
-\kappa_2^\prime\psi_2\overline{\psi}_2=0$. Hence, both $S$ and
$S^\prime$ also get heavy masses, and so $S=S^\prime =0$.

Similar to the original SUSY hybrid inflationary scenario,
inflation in this model can be initiated at a quasi-stable point
of
\begin{eqnarray} \label{initialCondi}
%
\Bigg \{
\begin{array}{l}
~(\kappa_1S+\kappa_1^\prime S^\prime )^2 ~\gtrsim ~
|\kappa_1^2M^2+\kappa_1^\prime\kappa_2^\prime M^{\prime 2}| ~,
\quad {\rm and}
\\
~(\kappa_2S+\kappa_2^\prime S^\prime )^2~\gtrsim ~
|\kappa_2^{\prime 2}M^{\prime 2}+\kappa_1\kappa_2 M^{2}| ~,
\end{array}
\end{eqnarray}
for which the tree level scalar potential is minimized at $\psi
_1=\overline{\psi}_1=\psi _2=\overline{\psi}_2=0$.
The left (right) hand sides of Eq.~(\ref{initialCondi}) come from
the (off-) diagonal components of the mass matrices for
$(\psi_1,\overline{\psi}_1^*)$ and $(\psi_2,\overline{\psi}_2^*)$.
Thus, inflation is described by the following effective
superpotential: 
\begin{eqnarray} \label{effsuperpot}
W=(\kappa S+\kappa'S')M^2 ,
\end{eqnarray}
where we redefined $\kappa$ and $\kappa^\prime$ as
\begin{eqnarray} \label{kappa2}
\kappa\equiv\kappa_1 \quad {\rm and}\quad \kappa^\prime\equiv
\kappa_2^\prime\left(\frac{M^{\prime 2}}{M^2}\right) .
\end{eqnarray}
We will assume a mild hierarchy between $\kappa$ and $\kappa'$,
i.e. $\kappa'/\kappa =(\kappa_2'M^{\prime 2}/\kappa_1M^2)\sim
{\cal O}(1-10^{-3})$, and $M\approx 4.5\times 10^{15}$ GeV. With
Eq.~(\ref{effsuperpot}), we obtain again the constant vacuum
energy at tree level, breaking SUSY. So the logarithmic quantum
corrections will be generated in the scalar potential as in the
single inflaton case.

The K${\rm \ddot{a}}$hler potential is expanded with the power of
$S^{(\prime)}/M_P$ ($\lesssim 1$) up to the quartic terms as
\begin{eqnarray} \label{Kahlerpot}
K=|S|^2+|S^\prime|^2+c_1\frac{|S|^4}{4M_P^2}+c_1^\prime\frac{
|S^\prime|^4}{4M_P^2} +c_2\frac{|S|^2|S'|^2}{M_P^2}
+\frac{c_3|S|^2+c_3^\prime |S^\prime|^2}{2M_P^2}(SS^{\prime
*}+S^*S^\prime) , \quad
\end{eqnarray}
where $c_i^{(\prime)}$ ($i=1,2,3$) are dimensionless coefficients.
The quartic terms' coefficients of order unity in the K${\rm
\ddot{a}}$hler potential would destroy the slow-roll condition of
the inflation. In the original version of the SUSY hybrid
inflation model, as mentioned in Introduction, the quartic term
coefficient $c_1$ in the K${\rm \ddot{a}}$hler potential,
$K=|S|^2+c_1|S|^4/4M_P^2+\cdots$, is assumed to be suppressed
($\lesssim 10^{-3} $) in order to satisfy Eq.~(\ref{ns}) as well
as the slow roll condition. Along the line of it, we also assume
one fine-tuned relation among the parameters of the K${\rm
\ddot{a}}$hler potential, $c_1$, $c_2$, and $c_3$:
\begin{eqnarray} \label{tuning}
c_1+\frac{c_3^2}{1-c_2} \equiv ~ \epsilon ~\lesssim ~ {\cal
O}(10^{-3}) .
\end{eqnarray}
Namely, $\epsilon$ is assumed to be of order $10^{-3}$ [Case ${\rm{\bf (A)}}$] or smaller [Case ${\rm{\bf (B,C)}}$]. 
Actually Eq.~(\ref{tuning}) can be satisfied, e.g. if they all are of order unity or
smaller, but related to each other by $c_1\approx
{-c_3^2}/{(1-c_2)}$ [Case ${\rm{\bf (A,B)}}$], or if they
(particularly $c_1$ and $c_3$) are sufficiently suppressed, $c_1$,
${c_3^2}/{(1-c_2)}\lesssim {\cal O}(10^{-3})$ [Case ${\rm{\bf
(A,C)}}$]. Indeed, $c_3$ can be made suppressed by introducing a
symmetry. We will propose later a simple idea, making $c_3$
adequately suppressed. In that case, only $c_1$ is assumed to be
small in the bare K${\rm \ddot{a}}$hler potential as in the
original version of the SUSY hybrid inflation.

With Eqs.~(\ref{effsuperpot}) and (\ref{Kahlerpot}), the
corrections coming from the scalar potential in SUGRA,
\begin{eqnarray} \label{sugrapot}
V_F=e^{K/M_P^2}\left[K_{ij^*}^{-1}D_iW(D_jW)^*-\frac{3}{M_P^2}|W|^2\right]
\end{eqnarray}
can be estimated. In our case, $i,j=\{S,S'\}$. $K_{ij^*}^{-1}$ and
$D_iW$ stand for the inverse K${\rm \ddot{a}}$hler metric and the
covariant derivative of the superpotential, respectively. Up to the
quadratic terms, their components are approximately given by
\begin{eqnarray}
&&\quad\quad ~~~ K_{SS^*}^{-1}\approx
1+\left(\frac{c_3^2}{1-c_2}-\epsilon\right)\frac{|S|^2}{M_P^2}-c_2\frac{|S^\prime|^2}{M_P^2}
-\frac{c_3}{M_P^2}(SS^{\prime *}+S^*S^\prime) ~, \label{Kss}
\\
&&\quad\quad\quad\quad\quad~ K_{S^\prime S^{\prime *}}^{-1}\approx
1-c_1^\prime\frac{|S^\prime|^2}{M_P^2}
-c_2\frac{|S|^2}{M_P^2}-\frac{c_3^\prime}{M_P^2}(SS^{\prime
*}+S^*S^\prime) ~,
\\
&&K_{SS^{\prime *}}^{-1}\approx -c_2\frac{S^\prime S^*}{M_P^2}
-c_3 \frac{|S|^2}{M_P^2} -c_3^\prime \frac{|S^\prime|^2}{M_P^2}
~,\quad
K_{S^{\prime}S^*}^{-1}\approx -c_2\frac{S S^{\prime *}}{M_P^2}
-c_3 \frac{|S|^2}{M_P^2}-c_3^\prime \frac{|S^\prime|^2}{M_P^2} ~,
\\
&&D_SW\approx M^2\left(\kappa+\kappa\frac{|S|^2}{M_P^2}+
{\kappa'}\frac{S^\prime S^*}{M_P^2}\right) ~, \quad
D_{S^\prime}W\approx M^2\left(\kappa^\prime
+\kappa^\prime\frac{|S^\prime|^2}{M_P^2}+ {\kappa}\frac{S
S^{\prime *}}{M_P^2}\right)  . \quad
\end{eqnarray}
In Eq.~(\ref{Kss}), we inserted Eq.~(\ref{tuning}). The scalar
potential Eq.~(\ref{sugrapot}) is, thus, estimated as
\begin{eqnarray} \label{scalarpot}
V_F&\approx&
\kappa^2M^4\left\{1+\left(\frac{c_3^2}{1-c_2}-\epsilon\right)|x|^2+(1-c_2)|y|^2-c_3(x^*y+xy^*)
\right\}\nonumber \\
&&+{\kappa^{\prime 2}}M^4\bigg\{1+(1-c_2)|x|^2-c_1^\prime
|y|^2-c_3^\prime (x^*y+xy^*)\bigg\}
\nonumber \\
&&-{\kappa^\prime}{\kappa}M^4 \bigg\{(1+c_2)(x^*y+xy^*)+2c_3
|x|^2+2c_3^\prime |y|^2\bigg\}
\nonumber\\
&=& (\kappa^2+\kappa^{'2})M^4+\kappa^2M^4\times (x^*~y^*){\cal M}
(x ~ y)^T ,
\end{eqnarray}
where $x\equiv S/M_P$, $y\equiv S'/M_P$, and the mass matrix
${\cal M}$ is given by
\begin{eqnarray}
{\cal M}=\left[
\begin{array}{cc}
\frac{c_3^2}{1-c_2}-\epsilon -\frac{\kappa'}{\kappa}\{2c_3
-\frac{\kappa'}{\kappa}(1-c_2)\} ~~&~~
-c_3-\frac{\kappa'}{\kappa}\left\{(1+c_2)+\frac{\kappa'}{\kappa}c_3^\prime\right\}
\\
-c_3-\frac{\kappa'}{\kappa}\left\{(1+c_2)+\frac{\kappa'}{\kappa}c_3^\prime\right\}
~~&~~ (1-c_2)-\frac{\kappa^{'}}{\kappa}\left\{
2c_3^\prime+\frac{\kappa'}{\kappa}~c_1^\prime\right\}
\end{array} \right] .
\end{eqnarray}

Case ${\rm{\bf (A)}}$: In the absence of the second inflatons's contribution to
the superpotential Eq.~(\ref{effsuperpot}), namely,
$\kappa'/\kappa\rightarrow 0$, the smaller eigenvalues of ${\cal
M}$ is given by
\begin{eqnarray}
-\epsilon\times\frac{(1-c_2)^2}{(1-c_2)^2+c_3^2} ~ ,
\end{eqnarray}
which can be identified with $\delta/2$ of Eq.~(\ref{toypot}).
Hence, if $\epsilon\sim {\cal O}(10^{-3})$, then the small
negative mass term of Eq.~(\ref{toypot}) can be supported purely
by the K${\rm \ddot{a}}$hler potential. Even if $c_1$ is quite
larger or smaller than ${\cal O}(10^{-3})$ and its sign is
positive, one can still obtain $n_s\approx 0.96$ by adjusting the
other parameters, $c_3^2$ and $c_2$. Particularly if $c_3\ll
c_2\sim {\cal O}(1)$, the smaller eigenvalue becomes just
$-\epsilon$. In this paper, however, we are more interested in the
case to acquire $n_s\approx 0.96$ with the help of the
{\it superpotential}, assuming $\epsilon \lesssim {\cal O}(10^{-3})$. It is because the superpotential is quantum
mechanically controllable unlike the K${\rm \ddot{a}}$hler potential.

Case ${\rm{\bf (B)}}$: If $c_1={-c_3^2}/{(1-c_2)}$ and
$\kappa'/\kappa \lesssim {\cal O}(1)$, the mass eigenstates and
eigenvalues during inflation are
\begin{eqnarray} \label{EvaluesB}
\left(
\begin{array}{c}\phi_L
\\
\phi_H
\end{array}\right)
\approx \frac{1}{D^{1/2}}\left[
\begin{array}{cc}
1-c_2 & c_3 \\
-c_3 & 1-c_2
\end{array} \right]
\left(
\begin{array}{c}
x\\
y~\end{array}\right)
\quad {\rm for}\quad
\Bigg\{
\begin{array}{l}
m_L^2 \approx
-\frac{\kappa'}{\kappa}\frac{2c_3(2-2c_2+c_3c_3')}{D} ~,
\\
m_H^2 \approx  1-c_2+\frac{c_3^2}{1-c_2} ~,
\end{array}
\end{eqnarray}
where $D\equiv (1-c_2)^2+c_3^2$.

Case ${\rm{\bf (C)}}$: If $c_1$, ${c_3^2}/{(1-c_2)} \ll {\cal
O}(1)$, fulfilling Eq.~(\ref{tuning}), and $\kappa'/\kappa
\lesssim {\cal O}(1)$, the mass eigenstates and eigenvalues are
given by
\begin{eqnarray} \label{EvaluesC}
\left(
\begin{array}{c}\phi_L
\\
\phi_H
\end{array}\right)
\approx \left[
\begin{array}{cc}
1 ~&~ \frac{\kappa'}{\kappa}\frac{1+c_2}{1-c_2} \\
-\frac{\kappa'}{\kappa}\frac{1+c_2}{1-c_2} ~&~ 1
\end{array} \right]
\left(
\begin{array}{c}
x\\
y~\end{array}\right)
\quad\quad {\rm for}\quad
\Bigg\{
\begin{array}{l}
m_L^2 ~\approx ~
-\left(\frac{\kappa'}{\kappa}\right)^2\frac{4c_2}{1-c_2} ~,
\\
m_H^2 ~\approx ~ 1-c_2 ~.
\end{array}
\end{eqnarray}
In both cases, the mass squared of the heavier component,
$\phi_H\times M_P$ is of the Hubble scale [$\sim {\cal
O}(\kappa^2M^4/M_P^2)$]. Consequently, it is expected to be stuck
to the origin during inflation, $\phi_H=0$. On the other hand, the
mass squared of the lighter component, $\phi_L\times M_P$ can be
much lighter than the Hubble scale, if $({\kappa'}/{\kappa})c_3$
for Case ${\rm{\bf (B)}}$ [or $({\kappa'}/{\kappa})^2c_2$ for Case
${\rm{\bf (C)}}$] is small enough. Therefore, inflation can be
driven only by lighter mass eigenstate. Moreover, the sign of
$m_L^2$ can be negative, if $c_3$ for Case ${\rm{\bf (B)}}$ [or
$c_2$ for Case ${\rm{\bf (C)}}$] is positive. $\phi_L$ can be
identified with the $\varphi$ of Eq.~(\ref{toypot}), and
$\mu^4=(\kappa^2+\kappa^{'2})M^4$. Quantum correction by the
coupling between $\phi_L$ and $\{\psi_{1,2},
\overline{\psi}_{1,2}\}$ would induce the logarithmic term in
Eq.~(\ref{toypot}), which leads $\phi_L$ eventually into the
origin. $\phi_L=\phi_H=0$ implies $S=S^\prime=0$. As $S$ and
$S^\prime$ approach the origin, Eq.~(\ref{initialCondi}) becomes
violated, and then $\psi_{1,2}$ and $\overline{\psi}_{1,2}$ also
roll down to the absolute minima, developing VEVs. Hence, SUSY is
recovered after inflation terminates.

Identification of $m_L^2$ with $\delta/2$ of Eq.~(\ref{ns}) yields
\begin{eqnarray} \label{sum}
\frac{\delta}{2}\approx -3.5\times 10^{-3} \approx \left\{
\begin{array}{l}
-\epsilon~\frac{(1-c_2)^2}{(1-c_2)^2+c_3^2} 
\quad\quad~~~~~ ~{\rm for}\quad {\rm Case}\quad {\rm{\bf (A)}} ~,
\\
-\frac{\kappa'}{\kappa}~\frac{2c_3(2-2c_2+c_3c_3')}{(1-c_2)^2+c_3^2}
\quad~~ ~{\rm for}\quad {\rm Case}\quad {\rm{\bf (B)}} ~,
\\
-\left(\frac{\kappa'}{\kappa}\right)^2\frac{4c_2}{1-c_2}
\quad\quad\quad~~~~ {\rm for}\quad {\rm Case}\quad {\rm{\bf (C)}} ~.~
\end{array}
\right.
\end{eqnarray}
In Case ${\rm{\bf (B)}}$, hence, ${\kappa'}/{\kappa}=
(\kappa_2^\prime M^{\prime 2})/(\kappa_1M^2)\sim {\cal
O}(10^{-3}-10^{-1})$ fulfills the constraint for $c_3\sim {\cal
O}(1-10^{-2})$.
Particularly, {\it if all the quartic terms's coefficients of the
K${\rm \ddot{a}}$hler potential are of order unity,
$\kappa'/\kappa$ should be of order ${\cal O}(10^{-3})$}.
On the other hand, {\it all the quartic terms in the K${\rm
\ddot{a}}$hler potential are suppressed, i.e. $c_i^{(\prime)}$
($i=1,2,3$) including $c_3$ are of order $10^{-2}$ or smaller,
$\kappa'/\kappa$ of order $10^{-1}$ is necessary}.

\begin{figure}[!t]
\begin{center}
   \includegraphics[width=0.5\textwidth]{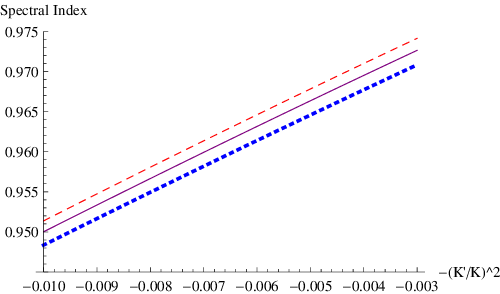}
\end{center}
\caption{Spectral index ($n_s$) vs. $\delta$.  
The three lines correspond to the cases of $N_e=60$ (red dashed line), $N_e=55$ (purple solid line), $N_e=50$ (blue dotted line). 
The spectral index is insensitive to ``$\varepsilon_s$'' i.e. the soft SUSY breaking ``A-term'' in the range of $0.94\lesssim n_s\lesssim 0.98$
Particularly, $\delta = -(\kappa^\prime/\kappa)^2$ for $4c_2/(1-c_2)=1/2$ in Case ${\rm{\bf (C)}}$.
}
\label{fig:n-delta}
\end{figure}
%

In Case ${\rm{\bf (C)}}$, ${\kappa'}/{\kappa}\sim {\cal
O}(10^{-2}-10^{-1})$ satisfies the constraint for $c_2\sim {\cal
O}(1-10^{-1})$.  Thus, the mildly hierarchical $\kappa'$ and
$\kappa$ couplings (or $\kappa_2^\prime M^{\prime 2}$ and
$\kappa_1M^2$) can generate the small negative inflaton's mass
squared, explaining $n_s\approx 0.96$.

\begin{figure}[!t]
\begin{center}
   \includegraphics[width=0.5\textwidth]{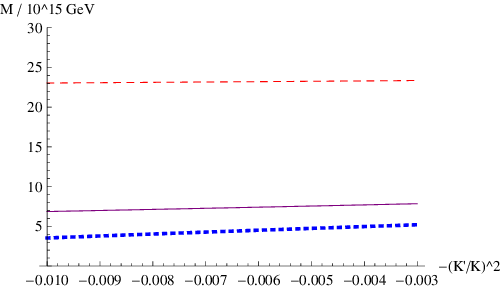}
\end{center}
\caption{Symmetry breaking scale ($M$) vs. $\delta$.
The three lines correspond to the cases of $\varepsilon_s/\sqrt{\alpha}=1.5$ (red dashed line), $\varepsilon_s/\sqrt{\alpha}=0.1$ (purple solid line), $\varepsilon_s/\sqrt{\alpha}=10^{-7}$ (blue dotted line). 
Particularly, $\delta= -(\kappa^\prime/\kappa)^2$ for $4c_2/(1-c_2)=1/2$ in Case ${\rm{\bf (C)}}$.}
\label{fig:M-delta}
\end{figure}

FIG. \ref{fig:n-delta} and \ref{fig:M-delta} show how $n_s$ and $M$ change as $\delta$  varied, based on the scalar potential
\begin{eqnarray} \label{Vanal}
V \approx \mu^4\left(1-\frac{\delta}{2}+\alpha ~{\rm log}~x+\frac{\delta}{2}x^2+\varepsilon_s x\right) .
\end{eqnarray}
Here we included the dominant soft SUSY breaking term, which is the ``A-term,'' $m_{3/2}\kappa SM^2$ [$=(\kappa^2 M^4)(m_{3/2}M_P/\kappa M^2)(S/M_P)\equiv\mu^4\varepsilon_s x$].\footnote{Since the quadratic term  coming from the non-minimal  K${\rm\ddot{a}}$hler potential in Eq.~(\ref{Vanal}) is much dominant over the soft mass term $m_{3/2}^2|S|^2$ unlike the case of Ref.~\cite{a-term}, we neglected the soft mass term of $S$ in Eq.~(\ref{Vanal}).} 
The presence of the ``$\varepsilon_s$ term'' in Eq.~(\ref{Vanal}) modifies $N_e$ of Eq.~(\ref{N,eta}) into 
\begin{eqnarray} \label{N2}
N_e=\frac{1}{2\delta}\left[{\rm log}\left(1+\frac{\delta}{\alpha}x^2+\frac{\varepsilon_s}{\alpha}x\right)+\frac{2\varepsilon_s}{\sqrt{-4\alpha\delta+\varepsilon_s^2}}{\rm Tanh}^{-1}\left(\frac{\varepsilon_s+2\delta x}{\sqrt{-4\alpha\delta+\varepsilon_s^2}}\right)\right] .
\end{eqnarray}
It corrects also ${\cal P}^{1/2}_{\cal R}$ of Eq.~(\ref{powerSP}) into 
\begin{equation} \label{pwSP2}
{\cal P}^{1/2}_{\cal R}=\sqrt{\frac23}\frac{M^2}{M_P^2}\left(\frac{x/\sqrt{\alpha}}{1+\delta x^2/\alpha+\epsilon x/\alpha}\right)\approx 4.93\times 10^{-5} ~. 
\end{equation}
But $\eta$ is still given by $\delta-\alpha/x^2$. It can be utilized to substitute $x$ with $\eta$ and $\delta$ in Eqs.~(\ref{N2}) and (\ref{pwSP2}). 
With Eqs.~(\ref{N2}), (\ref{pwSP2}), and $\eta=\delta-\alpha/x^2$, one can get the relations among $n_s$ ($\approx 1+2\eta$), $M$, and $\delta$ [or $\sim -(\kappa^\prime/\kappa)^2$], as seen in FIG. \ref{fig:n-delta} and \ref{fig:M-delta}.
We set $\varepsilon_s=m_{3/2}M_P/\kappa M^2=10^{-7}$ 
in FIG. \ref{fig:n-delta}.
The effect of the $\varepsilon_s$ term turns out to be almost negligible in the range of $0.94\lesssim n_s\lesssim 0.98$.
On the other hand, it can be important in determining  $M$, if $\alpha$ ($=\kappa^2/8\pi^2$) is small enough. 
As seen in Eq.~(\ref{sum}), $\delta/2$ is approximately $-(\kappa'/\kappa)^2[4c_2/(1-4c_2)]$ for Case ${\rm{\bf (C)}}$. For simplicity, we set $4c_2/(1-4c_2)=1/2$ in FIG. \ref{fig:n-delta} and \ref{fig:M-delta}. 
As discussed before, the center value of $n_s$ ($\approx 0.96$) in WMAP7 is possible when $\delta/2\approx -3.5\times 10^{-3}$.

%

Before finishing this section, let us briefly discuss how to
suppress $c_3$ and $\kappa^\prime$. Except the U(1)$_R$ symmetry,
there is another symmetry as mentioned above, under which 
$\psi_{1,2}$ and $\overline{\psi}_{1,2}$ in Eq.~(\ref{superPot})
can carry some proper charges. One can assign a charge also to
$S^\prime$ such that $S^\prime$ in the $c_3$ and $c_3^\prime$
terms of the K${\rm \ddot{a}}$hler potential (\ref{Kahlerpot})
and the $\kappa^\prime$ terms of the superpotential
(\ref{superPot}) should be replaced by $S^\prime\rightarrow
S^\prime(\psi_1/M_P)^n$, where $n$ is a positive integer
determined by the charge assignment to $S^\prime$. Note that such
a charge assignment for $S^\prime$ leaves intact the quadratic
terms and $c_1^{\prime}$, $c_2$ terms in the K${\rm \ddot{a}}$hler
potential.

If the non-renormalizable terms
$S(\psi_1\overline{\psi}_1)^2/M_P^2$ and
$S^\prime(\psi_1/M_P)^n(\psi_1\overline{\psi}_1)^2/M_P^2$ [and
also $S(\psi_2\overline{\psi}_2)^2/M_P^2$,
$S^\prime(\psi_1/M_P)^n(\psi_2\overline{\psi}_2)^2/M_P^2$]
are included in the superpotential (\ref{superPot}), a trajectory of
$\psi_1\overline{\psi}_1\lesssim {\cal{O}}(M_P^2)$ during (and
also after) inflation is allowed \cite{422}. Accordingly, $c_3$ in the K${\rm \ddot{a}}$hler potential and $\kappa^\prime$ in the
superpotential are effectively suppressed by the factor $(\psi_1/M_P)^n$.
Thus, one could assume that only $c_1$ is small in the {\it bare}
K${\rm \ddot{a}}$hler potential, satisfying Eq.~(\ref{tuning}).

\section{Two field Inflation}

In this section, we will discuss the possibility that the two
fields, i.e. $\phi_H$ and $\phi_L$ (or $S$ and $S^\prime$) both drive
inflation. Let us suppose that the mass of $\phi_H$ as well as
$\phi_L$ is lighter than the Hubble scale,
$(\kappa^2+\kappa^{\prime 2})^{1/2}M^2/M_P$.
Unless both of the leading terms for $\phi_L$ and $\phi_H$ in the scalar potential are logarithmic, 
then, the spectral index of $0.96$ would be easily achievable. 

Only with one inflaton field, CMB spectrum should be almost Gaussian, if the kinetic term of the inflaton is of the canonical form \cite{Maldacena}.
With two or more inflaton
fields \cite{LP,Naruko,KYC}, however, an observably large non-Gaussianity is allowed, because the mode of the isocurvature may cause it \cite{iso}. After the isocurvature modes are exhausted, the curvature perturbation is preserved during the radiation era.
In Refs.~\cite{Naruko,KYC}, large non-Gaussianity is discussed in the case that the leading terms of two inflatons in the scalar potential are all quadratic in hybrid inflation. In this case, a negative value of the non-linearity parameter $f_{\rm NL}$ is preferred, if it is large \cite{KYC}. 

In this section, we discuss the case that one of the leading scalar potential is given by a logarithmic term, while the other is quadratic:  
\begin{eqnarray} \label{scalarpot2}
V=\mu^4\left(1+\alpha{\rm log}\frac{\phi_l}{\Lambda}+\frac{\eta}{2}\frac{\phi_h^2}{M_P^2}\right) ~,
\end{eqnarray}
where $\alpha\approx\kappa^2/8\pi^2$ and $\eta\equiv M_P^2\partial_{\phi}^2V/V\sim {\cal O}(10^{-2})$. $\phi_l/M_P$ and $\phi_h/M_P$ are identified with $\phi_L$ and $\phi_H$, respectively, discussed in Sec. III. Particularly, $\phi_l=S$ and $\phi_h=S^\prime$ in Case (C). This scalar potential can be obtained e.g. if $\kappa^\prime/\kappa$  is quite small [or $(\kappa_2^\prime M^{\prime 2})/(\kappa_1M^2)\ll {\cal O}(1)$ in Eq.~(\ref{kappa2})], and $c_2$ is finely tuned [$1-c_2\sim{\cal O}(10^{-2})$] in Case (C).
The slow-roll parameters for the scalar potential Eq.~(\ref{scalarpot2}) are estimated as  
\begin{eqnarray} \label{sr2}
\epsilon_\chi=\frac{M_P^2}{2}\left(\frac{\alpha}{S}\right)^2\equiv ~\frac{\alpha}{2}\frac{1}{\chi^2} ~ &,&\quad \eta_\chi=-M_P^2\frac{\alpha}{S^2}\equiv -\frac{1}{\chi^2} ~, \\
\epsilon_\varphi=\frac{\eta^2}{2}\left(\frac{S^\prime}{M_P}\right)^2\equiv\frac{|\eta|}{2}\varphi^2 &,& \quad\quad\quad\quad \eta_\varphi=\eta ~, 
\end{eqnarray}
where we defined $\chi$ and $\varphi$ as $\chi\equiv S/\sqrt{\alpha}M_P$ and $\varphi\equiv\sqrt{|\eta|}S^\prime/M_P$, respectively.  50-60 e-folds ($=N_e$) of $\chi$ and $\varphi$ constrains the field values at the time of horizon exit 
and the end of inflation as follows:
\begin{eqnarray} \label{efolds2}
\chi_*^2 -\chi_e^2 =2 N_e ~, \quad\quad \varphi_*=e^{N_e\eta}\varphi_e ~,
\end{eqnarray}
where `$*$' and `$e$' denote the values evaluated at a few Hubble times after horizon exit and the end of inflation, respectively. 
Note that $|\eta_\chi^*|=1/\chi_*^2<1/(2 N_e)\approx 0.01$ for 50-60 e-folds.
In this paper, we focus on the case of  $\epsilon_\varphi^{*,e}\gg\epsilon_\chi^{*,e}$, since  the $\epsilon_\varphi^{*,e}\ll\epsilon_\chi^{*,e}$ case turns out not to yield large non-Gaussianity.  

With the $\delta N$ formalism \cite{deltaN}, the power spectrum and the spectral index can be  written in terms of the slow-roll parameters \cite{CHB,BCH}, 
\begin{eqnarray} \label{sI2}
{\cal P}_{\cal R}=\frac{\mu^4}{24\pi^2M_P^4\epsilon_\varphi^*}(1+\tilde{r}) ~, 
\quad\quad n_s-1= -2\epsilon^*+2\frac{-2\epsilon^*+\eta
+\tilde{r}\eta_\chi^*}{1+\tilde{r}} ~,  
\end{eqnarray}
where $\epsilon^*\equiv\epsilon^*_\chi+\epsilon^*_\varphi$,  and $\tilde{r}$ is defined as  
\begin{eqnarray} \label{sin}
\tilde{r}\equiv\frac{{\rm sin}^4\theta_e}{{\rm sin}^2\theta_*}~, \quad\quad {\rm and}\quad\quad {\rm sin}^2\theta_{e,*}\equiv\frac{\epsilon_\chi^{e,*}}{\epsilon_\varphi^{e,*} +\epsilon_\chi^{e,*}}\approx \frac{\epsilon_\chi^{e,*}}{\epsilon_\varphi^{e,*}}=\frac{\alpha/|\eta|}{\chi_{e,*}^2\varphi_{e,*}^2} ~. 
\end{eqnarray}
The power spectrum in Eq.~(\ref{sI2}) determines the Hubble scale, $\mu^2/M_P$ ($=|\kappa| M^2/M_P$). 
If $\tilde{r}\gg 1$, the spectral index in Eq.~(\ref{sI2}) reduces  to $n_s\approx 1+2\eta^*_\chi\approx 1-\frac{1}{N_e}\approx 0.98$ from Eqs.~(\ref{sr2}) and  (\ref{efolds2}). 
In order to get $n_s\approx 0.96$, hence, we should take $\tilde{r}\ll 1$ or $\tilde{r}\sim {\cal O}(1)$. Then, $n_s$ is contributed mainly by $\eta$, and so it should be a negative value. 
Since $n_s-1\approx-0.04$ and $\epsilon^*\ll\eta$, $\eta_\chi^*$ ($=-1/\chi_*^2$), the second equation in Eq.~(\ref{sI2}) can be simplified as  
\begin{equation} \label{simpleta}
\eta \approx -0.02(1+\tilde{r})+\frac{\tilde{r}}{\chi_*^2} ~.
\end{equation}
$\eta<0$ implies that the field value of $\phi_h$ [or $S^\prime$ in Case (C)] should increase during inflation from Eq.~(\ref{efolds2}) [while $\phi_l$ (or $S$) decreases]. As will be seen, $S^\prime_e/S^\prime_*=\varphi_e/\varphi_*\sim 3$ and $S_e/S_*=\chi_e/\chi_*\sim 0.1$ for  the cases in which we are interested.

We suppose that the first condition of Eq.~(\ref{initialCondi}) is first violated. It can be easily achieved e.g. if $\kappa_2\approx 0$ and $M^\prime\ll M$, because $S^\prime$ increases during inflation. 
Recall that the first (second) inequality in Eq.~(\ref{initialCondi}) is the condition of the positive mass squared for the waterfall fields $\psi_1$ and $\overline{\psi}_1$ ($\psi_2$ and $\overline{\psi}_2$) at the origin.
Once the first condition of Eq.~(\ref{initialCondi}) violated, thus, the waterfall fields $\psi_1$ and $\overline{\psi}_1$ become tachyonic at the origin. Hence, $\psi_1$ and $\overline{\psi}_1$ should develop non-zero VEVs at their local minima.


Assuming $\psi_1$ and $\overline{\psi}_1$ develop VEVs in the real direction, the D-flat condition $\psi_1=\overline{\psi}_1^*$ ($\equiv\psi_r/\sqrt{2}$) yields the following dominant scalar potential: 
\begin{eqnarray} \label{endPot}
V=\left(\kappa^2+\kappa^{\prime 2}\right)M^{4}-\left[\kappa^2M^2+\kappa_1^\prime\kappa_2^{\prime}M^{\prime 2}-\left|\kappa S+\kappa_1^{\prime}S^{\prime}\right|^2\right]\psi_r^2+(\kappa^2+\kappa_1^{\prime 2})\frac{\psi_r^4}{4} ~
\end{eqnarray}
from the superpotential (\ref{superPot}), because still  $\langle\psi_2\rangle=\langle\overline{\psi}_2\rangle=0$.
Here we dropped the mass term of $S^\prime$ coming from the K${\rm\ddot{a}}$hler potential; it is much smaller than $\kappa_1^{\prime 2}\psi_r^2|S^{\prime}|^2$ in Eq.~(\ref{endPot}), if $\kappa_1^{\prime 2}/(\kappa_1^{\prime 2}+\kappa^2)\gg\eta$ or $\kappa_1^{\prime 2}\gg\kappa^2$. It is because $\psi_r^2$ turns out to be of order $V/M_P^2(\kappa^2+\kappa_1^{\prime 2})$, as will be seen below.  
Note that the heavier mass eigenstate is $(\kappa S+\kappa_1^\prime S^\prime)/\sqrt{\kappa^2+\kappa_1^{\prime 2}}$, 
when $\langle\psi_r\rangle\neq 0$.
The VEV of $\psi_r$ can be determined at its local minimum, satisfying $\partial_{\psi_r}V=0$:  
\begin{eqnarray} \label{psi}
\frac12(\kappa^2+\kappa_1^{\prime 2})\psi_r^2=\kappa^2M^2+\kappa_1^{\prime}\kappa_2^{\prime}M^{\prime 2}-\left|\kappa S+\kappa_1^{\prime}S^{\prime}\right|^2 ~.
\end{eqnarray} 
        
Inflation terminates when the heavier mass eigenstate, $(\kappa S+\kappa_1^\prime S^\prime)/\sqrt{\kappa^2+\kappa_1^{\prime 2}}$ ($\equiv \tilde{S}$) violates the slow roll condition: 
$\eta_{\tilde{s}}=M_P^2\partial_{\tilde{s}}^2V/V\sim{\cal O}(1)$, that is to say, 
\begin{equation} \label{endcondi}
M_P^2(\kappa^2+\kappa_1^{\prime 2})\psi_r^2= cV ~,
\end{equation}
where $c$ is a number of order unity ($\approx 3$ \cite{Kolb}). It is the condition for the end of inflation \cite{NHinf}.   
With Eqs.~(\ref{endPot}) and (\ref{psi}), the condition for the end of inflation, Eq.~(\ref{endcondi}) 
becomes    
\begin{equation}
V=\frac{2}{c^2}(\kappa^2+\kappa_1^{\prime 2})M_P^4\left\{-1+\sqrt{1+\frac{c^2(\kappa^2+\kappa^{\prime 2})M^{4}}{(\kappa^2+\kappa_1^{\prime 2})M_P^4}}\right\} \approx (\kappa^2+\kappa^{\prime 2})M^{4} ~. 
\end{equation}
Namely, the condition for end of inflation implies 
$V(S,S^\prime)=$ constant, which is independent of the exact value of $c$. 
Since the condition for end of inflation is given by a uniform energy density condition, the end point effect of non-Gaussianity would be negligible \cite{NHinf,CKK}.\footnote{The end point effect in non-Gaussianity for the case that two inflaton masses are hierarchical ($m_1^2\gg m_2^2$) has been discussed in Ref.~\cite{Naruko}. As will be seen later, 
$m_1^2$ ($=\eta_\chi$) and $m_2^2$ ($=\eta$) in our model are estimated as $|\eta_\chi|\sim {\cal O}(1)\gg |\eta|\sim {\cal O}(10^{-2})$, when inflation is over. 
Following the notation of Ref.~\cite{Naruko}, $G=2|\kappa S+\kappa_1^\prime S^\prime|^2$ and  $\lambda=(\kappa^2+\kappa_1^{\prime 2})$. Hence,  $g_1^2=0$, $g_2^2=2(\kappa^2+\kappa_1^{\prime 2})$, and  ${\rm sin}\alpha=-\kappa/\sqrt{\kappa^2+\kappa_1^{\prime 2}}$, ${\rm cos}\alpha=\kappa_1^\prime/
\sqrt{\kappa^2+\kappa_1^{\prime 2}}$. Since $\kappa_1^{\prime 2}\gg \kappa^2$ in our case, $\alpha\approx 0$. Our model corresponds to the case of $\beta=\gamma=\pi/2$ and $\delta=0$ of Ref.~\cite{Naruko}. With such parameters, the end point effect in non-Gaussianity cannot be large. } 

After the heavier mass eigenstate, $\tilde{S}$ violates the slow roll condition, it becomes more accelerated toward the origin, and at some point it violates also the second condition of Eq.~(\ref{initialCondi}). So $\psi_2$ and $\overline{\psi}_2$ also roll down to their true minima, developing large VEVs. 
Hence, $S$ and $S^\prime$ both eventually return to their SUSY minima. 


In the limit of ${\rm sin}^2\theta_*\ll{\rm sin}^2\theta_e\ll 1$, the non-linearity parameter $f_{\rm NL}$ can be potentially large \cite{BCH}. In this limit, $f_{\rm NL}$ approximately becomes\footnote{
In fact, the K${\rm\ddot{a}}$hler potential in this case is not minimal, as seen in Eq.~(\ref{Kahlerpot}). Hence, the kinetic terms of the inflaton fields are not of the canonical form.  
However, the non-canonical effects \cite{CHB} give just the correction of $f_{\rm NL}\times {\cal O}(S^{(\prime)2}/M_P^2\ll 1)$ to Eq.~(\ref{fNL}), even if the coefficients in the K${\rm\ddot{a}}$hler potential were of order unity.
}  
\begin{eqnarray} \label{fNL}
f_{\rm NL}\approx\frac{5~\tilde{r}^2}{6~{\rm sin}^2\theta_e(1+\tilde{r})^2}
\left(-\eta^*_\chi+2\eta^e_\chi\right) ~.
\end{eqnarray}
Since $|\eta^*_\chi|\ll |\eta^e_\chi|$ and $\eta^*_\chi<0$, 
$f_{\rm NL}$ in this case should be negative, if it is large. With ${\rm sin}^2\theta_e=\tilde{r}~{\rm sin}^2\theta_*/{\rm sin}^2\theta_e=
\tilde{r}~\chi_e^2\varphi_e^2/\chi_*^2\varphi_*^2=
\tilde{r}~\chi_e^2e^{-2N_e\eta}/\chi_*^2$ 
from Eqs.~(\ref{sin}) and (\ref{efolds2}) and using Eqs.~(\ref{efolds2}), (\ref{sr2}), and (\ref{simpleta}), Eq.~(\ref{fNL}) becomes
\begin{equation} \label{fNL2}
-f_{\rm NL}\approx \left(\frac{5\chi_*^2}{3\chi_e^4}\right)\frac{\tilde{r}}{(1+\tilde{r})^2}~e^{2N_e\eta}\approx \frac{5(\chi_e^2+2N_e)}{3\chi_e^4}~\frac{\tilde{r}}{(1+\tilde{r})^2}~
e^{-2N_e[0.02(1+\tilde{r})-\tilde{r}/(\chi_e^2+2N_e)]} ~.
\end{equation}
Since the constraint by the WMAP7 data is $-10<f_{\rm NL}^{\rm local}<74$ in the $95\%$ confidence level, the absolute value of $f_{\rm NL}$ should be smaller than $10$. In TABLE I, we present the some parameters yielding negative non-Gaussianity of order unity and $n_s=0.96$. 
%

For given $N_e$, $\chi_e$ and $\tilde{r}$,  $f_{\rm NL}$ and $\eta$ are determined from Eqs.~(\ref{fNL2}) and (\ref{simpleta}). Then, ${\rm sin}^2\theta_e$ and ${\rm sin}^2\theta_*$ can be also estimated from the relations, ${\rm sin}^2\theta_e=\tilde{r}\chi_e^2e^{-2N_e\eta}/\chi_*^2$ and ${\rm sin}^2\theta_*={\rm sin}^4\theta_e/\tilde{r}$. Note that $f_{\rm NL}$, $\eta$, ${\rm sin}^2\theta_e$, ${\rm sin}^2\theta_*$, etc. do not depend on the each value of $\varphi_e$ or $\varphi_*$ but on the ratio $\varphi_e/\varphi_*$ ($=e^{-N_e\eta}\sim 3$): $\varphi_e$ or $\varphi_*$ is necessary only to determine $\alpha$ via  Eq.~(\ref{sin}). $\alpha$ can be small enough by taking small $\varphi_e$ or $\varphi_*$.     

\begin{table}
\begin{tabular}{c||cccccc}
\hline  
~ $f_{\rm NL}$ ~&~ $N_e$~&~ $\chi_e$ ~& $\tilde{r}$ & $\eta$ & ${\rm sin}^2\theta_e$ & ${\rm sin}^2\theta_*$ 
\\
\hline 
$-10$ &~ $55$ ~& $0.7$ &~ $0.245$ ~& $-0.023$ &~ $1.3\cdot 10^{-2}$ ~&~ $7.1\cdot 10^{-4}$
\\
$-5$ &~ $55$ ~& $0.8$ &~ $0.167$ ~& $-0.022$ &~ $1.1\cdot 10^{-2}$ ~&~ $6.8\cdot 10^{-4}$
\\
$-3$ &~ $50$ ~& $1.0$ &~ $0.306$ ~& $-0.023$ &~ $3.0\cdot 10^{-2}$ ~&~ $3.0\cdot 10^{-3}$
\\
$-1$ &~ $60$ ~& $1.2$ &~ $0.255$ ~& $-0.023$ &~ $4.8\cdot 10^{-2}$ ~&~ $1.1\cdot 10^{-3}$
\\
\hline
\end{tabular}
\caption{Some parameter values, yielding $n_s=0.96$. The values of ${\rm sin}^2\theta_{e}$ and ${\rm sin}^2\theta_{*}$ listed in the table fulfill ${\rm sin}^2\theta_*\ll{\rm sin}^2\theta_e\ll 1$, ensuring the validity of Eq.~(\ref{fNL}). 
}
\label{tb:contents}
\end{table}


\section{Conclusion}

In this paper, we have studied the SUSY hybrid inflation in the
case that one more singlet field carrying the same quantum number
with the inflaton is present.
A parameter space of the K${\rm \ddot{a}}$hler potential
[$c_1+c_3^2/(1-c_2)\sim {\cal O}(10^{-3})$] permits $n_s\approx
0.96$. However, the K${\rm \ddot{a}}$hler potential is not
quantum mechanically controllable. It was pointed out that by such
one more singlet, the superpotential can mainly control the spectral index to $n_s\approx
0.96$, once one fine-tuning in the K${\rm \ddot{a}}$hler potential is assumed.

In this model, inflation is dominated by the superpotential,
$W=(\kappa S+\kappa' S')M^2$, but only one linear combination of
$S$ and $S'$ drives inflation. The smallness of $\kappa'/\kappa$
[$\sim {\cal O}(10^{-1}$--$10^{-3})$] in the superpotential can be responsible for the small negative mass squared of the
inflaton needed for explaining $n_s\approx 0.96$.

We also discuss non-Gaussianity, when the both singlets become light enough, and one singlet is governed by a logarithmic potential and the other by a quadratic one.
In this case, the non-linearity parameter $f_{\rm NL}$ is constrained to a negative value, if it is large. 

\acknowledgments{ \noindent The author thanks 
Ki-Young Choi for valuable discussions. This research is supported
by Basic Science Research Program through the National Research
Foundation of Korea (NRF) funded by the Ministry of Education,
Science and Technology (Grant No. 2010-0009021), and also by Pusan National University Research Grant, 2010. }


\end{document}